\documentclass[citesort,twocolumn,aps,prl,draft,showpacs,amssymb,floatfix]{revtex4}

\usepackage{graphicx}
\usepackage{bm}

\newcommand{\dd}{\mbox{d}}

\begin{document}

\title{Multifractal statistics of Lagrangian velocity and
acceleration in turbulence}

\author{L.~Biferale$^1$, G.~Boffetta$^{2,6}$, A.~Celani$^3$, 
B.J. Devenish$^1$, A.~Lanotte$^4$, and F.~Toschi$^5$}

\affiliation{
$^1$ Dipartimento di Fisica and INFM,
Universit\`a degli Studi di Roma ``Tor Vergata'', 
Via della Ricerca Scientifica 1, 00133 Roma, Italy \\
$^2$ Dipartimento di Fisica Generale and INFM,
Universit\`a degli Studi di Torino, 
Via Pietro Giuria 1, 10125, Torino, Italy\\
$^3$ CNRS, INLN, 1361 Route des Lucioles, 06560 Valbonne,
France \\
$^4$ CNR-ISAC, Sezione di Lecce, 
Str. Prov. Lecce-Monteroni km.1200, 73100 Lecce, Italy\\
$^5$ Istituto per le Applicazioni del Calcolo, CNR,
Viale del Policlinico 137, 00161 Roma, Italy\\
$^6$ CNR-ISAC, Sezione di Torino, Corso Fiume 4, 10133 Torino, Italy }
\date{\today}

\begin{abstract}
The statistical properties of velocity and acceleration fields along the trajectories
of fluid particles transported by a fully developed turbulent flow are
investigated by means of high resolution direct numerical simulations.
We present results for Lagrangian velocity structure functions, the 
acceleration probability density function and the acceleration variance conditioned on the instantaneous
velocity. These are compared with predictions of the multifractal formalism and
its merits and limitations  are discussed.
\end{abstract}


\maketitle
Understanding the  Lagrangian statistics of particles advected
by a turbulent velocity field, $\bm u(\bm x,t)$, is important 
both for its theoretical implications \cite{K65} and for applications, such
as the development of phenomenological and stochastic models 
for turbulent mixing \cite{pope}. Recently, several authors have attempted 
to describe Lagrangian statistics such as acceleration
by constructing models based on equilibrium statistics
(see e.g. \cite{beck,aringazin,arimitsu}, critically reviewed in \cite{GK03}). 
In this letter we show how the multifractal formalism 
offers an alternative approach which is rooted in the phenomenology of turbulence.
Here, we derive the Lagrangian statistics from the Eulerian statistics
without introducing {\em ad hoc} hypotheses.

In order to obtain an accurate description of the particle statistics
it is necessary to measure the positions, $\bm X(t)$, and velocities,
$\bm v(t) \equiv \dot{\bf X}(t) = \bm u({\bf X}(t),t)$, of the
particles with very high resolution, ranging from fractions of the
Kolmogorov timescale, $\tau_{\eta}$, to multiples of the Lagrangian
integral time scale, $T_L$.  The ratio of these timescales,
$T_L/\tau_{\eta}$, gives an estimate of the micro-scale Reynolds
number, $R_\lambda$, which may easily reach values of order $10^3$ in
laboratory experiments.  Despite recent advances in experimental
techniques for measuring Lagrangian turbulent statistics
\cite{cornell,pinton,ott_mann}, direct numerical simulations (DNS)
still offer higher accuracy albeit at a slightly lower Reynolds number
\cite{yeung,BS02,IK02,GF01}. 
In this letter we are concerned with single particle
statistics, that is, the statistics of velocity and acceleration
fluctuations along individual particle trajectories.  Here, we analyse
Lagrangian data obtained from a recent DNS of homogeneous isotropic
turbulence \cite{biferale04} which was performed on $512^3$ and
$1024^3$ cubic lattices with Reynolds numbers up to $R_\lambda \sim
280$. The Navier-Stokes equations were integrated using fully
de-aliased pseudo-spectral methods for a total time $T \approx T_L$.
Millions of Lagrangian particles (passive tracers) were released into
the flow once a statistically stationary velocity field had been
obtained. The positions and velocities of the particles were stored at
a sampling rate of $0.07 \tau_\eta$. The Lagrangian velocity was
calculated using linear interpolation.
Acceleration was calculated both by following the particle
and by direct computation from  all three forces acting on the particle
-- the pressure gradients, viscous forces and the large scale forcing.
The two measurements were found to be in very good agreement.
The flow was forced by keeping the total energy constant in 
the first two wavenumber shells (for more details see \cite{biferale04}).

It is well known that Lagrangian velocity increments, $\delta_\tau \bm
v = \bm v(t+\tau)-\bm v(t)$, are quasi-Gaussian for time lags $\tau$
of order $T_L$ but become increasingly intermittent at higher
frequencies \cite{pinton}. The resulting acceleration statistics
exhibit some of the most extreme fluctuations of any known quantity,
with accelerations, $\bm a (t)$, up to 80 times its root mean square
value, $a_{rms}$, possible
\cite{cornell}. The most natural way to quantify such phenomena is via
 probability density functions (pdfs) of the Lagrangian velocity
increment, ${\cal P}(\delta_\tau \bm v)$, and acceleration, ${\cal
P}(\bm a)$.  The frequency of extreme events is
reflected in the size of the tails of the pdfs and thus in the high
order moments. These can be analysed with the aid of Lagrangian
velocity structure functions $S_p(\tau) = \langle (\delta_\tau v)^p
\rangle$ where $\delta_{\tau} v$ characterises the magnitude of a
component of the velocity increment. Since the flow here is isotropic
the choice of component is immaterial.

It has long been recognised that Eulerian velocity fluctuations
in the inertial subrange  exhibit anomalous scaling: 
$\langle (\delta_r u)^p \rangle \equiv 
\langle (u(x+r)-u(x))^p \rangle\sim r^{\zeta_E(p)}$ \cite{frisch},
where $r$ is the spatial separation.  
On the basis of simple phenomenological arguments, 
we may expect the Lagrangian velocity fluctuations to exhibit
a power law behaviour for time scales within the inertial subrange too.
We may therefore assume that $ S_p(\tau) \sim \tau^{\zeta_L(p)}$ 
with $\tau_{\eta} \ll \tau \ll T_L$.
 Anomalous scaling is often 
interpreted as the result of the intermittent nature of the energy cascade.
Among the simplest stochastic models able to reproduce both qualitatively and 
quantitatively such intermittency are those based
on the multifractal formalism. 
This has been successfully used to explain Eulerian statistics such as 
structure functions \cite{frisch,parisi_frisch,benzi84} and velocity gradients 
\cite{ben91,biferale93}
and Lagrangian statistics such as the acceleration covariance
\cite{borgas93} and the velocity statistics \cite{chevillard,BDM02}.
The aim of this letter is to compare predictions of the
multifractal formalism for Lagrangian velocity structure functions,
the acceleration pdf and the acceleration variance conditioned on the
instantaneous velocity with those obtained from the DNS data. The
Lagrangian multifractal predictions are derived from the multifractal
formalism in the Eulerian reference frame without any additional free
parameters. 

In the multifractal formalism the global scale invariance
of Kolmogorov's theory (K41) becomes a local
scale invariance. Namely, the turbulent flow is assumed to possess a 
range of scaling exponents $I=(h_{\min},h_{\max})$. For each $h \in I$ there 
is a set $S_h \in {\mathbb R}^3$ of fractal dimension $D(h)$ such that, in the limit
of small $r$: $ \delta_r u(\bm x) \sim u_0 (r/L_0)^{h(\bm x)}$
for $ \bm x \in S_h$. Here $u_0$ is the large scale fluctuating velocity and
$L_0$ is the integral length scale. For small values of $u_0$ we are in the 
laminar part of the flow for which a multifractal description is not appropriate. 
From this local scaling law, the scaling properties of the 
Eulerian structure function can easily be derived by integrating over 
all possible $h$ \cite{frisch}:
$\langle (\delta_r u)^p \rangle \sim \langle u_0^p \rangle \int_I dh (r/L_0)^{hp + 3 - D(h)}$.
The factor $(r/L_0)^{3-D(h)}$ is the probability of being 
within a distance of order 
$r$ of the set $S_h$ of dimension $D(h)$. 
A saddle point approximation in the limit 
$r\ll L_0$ then gives the scaling exponents 
\begin{equation}
\zeta_E(p) =\inf_h(hp + 3 - D(h)).
\label{eq:zeta_e}
\end{equation}
If the Eulerian scaling exponents are known, $D(h)$ can be calculated from the
inverse of the Legendre transformation (\ref{eq:zeta_e}). 
Among possible empirical formulae for the scaling exponents, $\zeta_E(p)$, 
we choose the one of She and L\'ev\^eque \cite{she_leveque}.
Using this it can be shown that
\begin{equation}
 D(h)=1 + p^*(h) \left( h - \frac{1}{9} \right) + 
2 \left( \frac{2}{3} \right)^{\frac{p^*(h)}{3}}, 
\label{eq:d_h}
\end{equation}
where $p^*(h)=(3/\ln(2/3)) \ln[(1-9h)/(6 \ln(2/3))]$ is the value of $p$ 
which minimises the inverse of (\ref{eq:zeta_e}).

The velocity fluctuations along a particle trajectory may be considered 
as the superposition of different contributions from eddies
of all sizes. In a time lag $\tau$ the contributions from eddies
smaller than a given scale, $r$, are uncorrelated and one 
may then write $\delta_{\tau} v \sim \delta_r u$. We assume that  
$r$ and $\tau$ are linked by the typical eddy turn over time at 
the given spatial scale, $\tau_r \sim r/\delta _r u $.
 Therefore, in the multifractal terminology, 
\begin{equation}
\label{eq:tau_l}
\delta_{\tau} v \sim \delta_r u 
\qquad \tau \sim  \frac{L^h_0}{v_0} r^{1-h}. 
\end{equation}
The presence of fluctuating eddy turn over times is the only 
additional complication introduced by the multifractal formalism in the 
Lagrangian reference frame. Using (\ref{eq:tau_l}) we can now 
derive a prediction for the Lagrangian velocity structure function 
\cite{BDM02}:
$$
S_p(\tau) \sim \langle v_0^p \rangle 
\int_{h \in I} \dd h \left( \frac{\tau}{T_L} \right)^{\frac{hp + 3-D(h)}{1-h}},
$$
where the factor $(\tau/T_L)^{(3-D(h))/(1-h)}$ 
is the probability of observing an exponent $h$  in  a time lag 
$\tau$. The exponents $\zeta_L(p)$ then follow 
from a saddle point approximation in the limit $\tau  \ll T_L$:
\begin{equation}
\zeta_L(p) = \inf_h \left( \frac{hp + 3-D(h)}{1-h} \right).
\label{eq:zeta_l}
\end{equation}

%%%%%%%%%%%%%%%%%%%%%%%%%%%%%%%%%%%%%%%%%%%%%%%%%%%%%%%%%%%%%%%%
\begin{figure}[h]
\includegraphics[draft=false,scale=0.6]{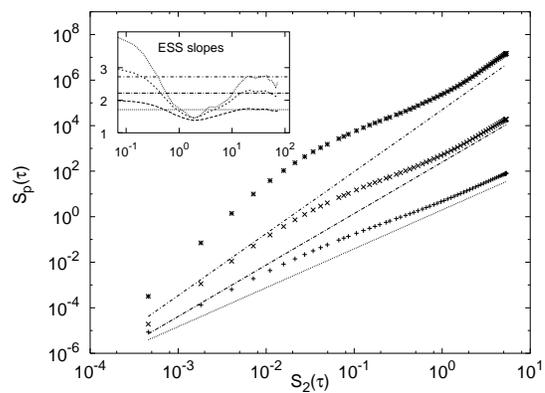}
\caption{ESS plot of  Lagrangian velocity structure 
function $S_p(\tau)$ versus $S_2(\tau)$.
Symbols refer to the DNS data for $p=8,6,4$ from top to bottom.
Lines have slopes $\zeta_L(p)/\zeta_L(2)$ given by the multifractal
prediction (\ref{eq:zeta_l}). In the inset we show the local slopes of
the DNS data and the multifractal predictions versus the time lag, $\tau$.}
\label{fig:vel}
\end{figure}
%%%%%%%%%%%%%%%%%%%%%%%%%%%%%%%%%%%%%%%%%%%%%%%%%%%%%%%%%%%%%%%%
In Fig.~(\ref{fig:vel}) the results for $S_p(\tau)$ calculated from
the DNS are presented. Although the scaling in a log-log plot is
reasonable, a more detailed inspection of logarithmic local slopes
displays a deterioration of scaling properties at small times. This is
due to the presence of a strong bottleneck for time lags, $\tau \in
[\tau_\eta, 10\tau_\eta]$.  This has been explained in terms of
trapping events inside vortical structures
\cite{biferale04}, a dynamical 
effect which may strongly affect scaling properties and which a simple
multifractal model cannot capture. For this reason, scaling properties are recovered only using Extended Self Similarity (ESS) \cite{benzi93} 
and for large time lags, $\tau > 10 \tau_\eta$.
 In this
interval a satisfactory agreement with the multifractal prediction
(\ref{eq:zeta_l}) is observed, namely $\zeta_L(4)/\zeta_L(2) = 1.71;
\zeta_L(6)/\zeta_L(2)=2.16;
\zeta_L(8)/\zeta_L(2)=2.72$. 

Similar phenomenological arguments can be used to
derive predictions for the acceleration statistics. 
The acceleration at the smallest scales is defined by 
\begin{equation}
a \equiv \frac{\delta_{\tau_\eta} v}{\tau_\eta}.
\label{eq:a}
\end{equation}
As the Kolmogorov scale, $\eta$, fluctuates in the multifractal 
formalism \cite{frisch}:
$
\eta(h,v_0) \sim \left(\nu L_0^h/v_0 \right)^{1/(1 + h)},
$
so does the Kolmogorov time scale, $\tau_\eta(h,v_0)$. Using 
(\ref{eq:tau_l}) and (\ref{eq:a}) evaluated 
 at  $\eta$, we get, for a given $h$ and $v_0$,
\begin{equation}
a(h,v_0) \sim \nu^{\frac{2h - 1}{1 + h}} v_0^{\frac{3}{1 + h}} L_0^{-\frac{3h}{1 + h}}.
\label{acc}
\end{equation}
The pdf of the acceleration can be derived by integrating (\ref{acc})
over all $h$ and $v_0$, weighted with their respective probabilities,
$(\tau_{\eta}(h,v_0)/T_L(v_0))^{(3-D(h))/(1-h)}$ and ${\cal P}(v_0)$.
The large scale velocity pdf is reasonably approximated by a Gaussian \cite{frisch}:
${\cal P}(v_0) = \exp(-v_0^2/2\sigma^2_v)/\sqrt{2\pi \sigma^2_v}$ 
, where $\sigma^2_v = \left< v_0^2 \right>$. Integration over 
$v_0$ gives:
\begin{eqnarray}
&{\cal P}(a) \sim & \int_{h\in I}  \dd h \,
a^{\frac{h-5+D(h)}{3}} \nu^{\frac{7-2h-2D(h)}{3}} L_0^{D(h)+h-3} \sigma_v^{-1} \times \nonumber \\
&& \quad 
\exp \left( -\frac{a^{\frac{2(1+h)}{3}} \nu^{\frac{2(1-2h)}{3}} L_0^{2h} }{2 \sigma^2_v} \right).
\label{eq:pdf_acc}
\end{eqnarray}

From (\ref{eq:pdf_acc}) we can derive the Reynolds number dependence of 
the acceleration moments \cite{borgas93,sawford}. For example,
in the limit of large $R_{\lambda}$ the second
order moment is given by $\left< a^2 \right> \, \propto  
R_{\lambda}^{\chi}$ where
$\chi = \sup_h \left( 2 (D(h)-4h-1)/(1+h) \right)$. Thus, we find that
$\chi=1.14$ which differs slightly 
from the K41 scaling, $\chi^{K41}=1$
 (see \cite{sawford,hill} for a discussion on departures
from K41 scalings in the context of acceleration statistics). 
In order to compare the DNS data with the multifractal 
prediction we normalise the acceleration by the rms acceleration, 
$\sigma_a=\langle a^2\rangle^{1/2}$. In terms of the
dimensionless acceleration, $\tilde a = a/\sigma_a$, (\ref{eq:pdf_acc}) becomes
\begin{equation}
{\cal P}(\tilde{a}) \sim
\int _{h \in I}\tilde{a}^{\frac{(h-5+D(h))}{3}} R_{\lambda}^{y(h)}
\exp \! \left( - \frac{1}{2}\tilde{a}^{\frac{2(1+h)}{3}} 
R_{\lambda}^{z(h)} \right) \, \dd h,
\label{pdf_norm}
\end{equation}
where $y(h) = \chi (h-5+D(h))/6 + 2(2D(h)+2h-7)/3$
and $z(h) = \chi (1+h)/3 + 4(2h-1)/3$.  We note
that (\ref{pdf_norm}) may show an unphysical divergence for
$a \approx 0$ for many multifractal models of $D(h)$.  For example,
with $D(h)$ given by (\ref{eq:d_h}) we cannot normalise 
${\cal P}(a)$ for $h <h_c \approx 0.16$.  This shortcoming is 
unimportant
for two reasons. First, as already stated, the multifractal formalism cannot be
trusted for small velocity and acceleration increments because it is 
based on arguments valid only to within a constant of order one. 
Thus, it is not suited for predicting precise
functional forms for the core of the pdf. Second, values of
$h \lesssim h_c$ correspond to very intense velocity 
fluctuations which have never been accurately tested in experiments or by  
DNS. The precise functional form of $D(h)$ 
for those values  of $h$ is therefore unknown. Thus, we restrict 
$h$ to be in the range $h_c<h \le h_{\max}$.  
For $h_{\max}$ we take the value of $h$ which satisfies $D'(h)=0$, that is
$h_{\max} \approx 0.38$.  Values of $h>h_{\max}$ affect only the peak
of the velocity distribution which we have 
 already excluded from our discussion.
 We also restrict  $|{\tilde a}|$ to lie in the
range $[\tilde a_{\min}, \infty)$ with $\tilde a_{\min}=O(1)$.

%%%%%%%%%%%%%%%%%%%%a%%%%%%%%%%%%%%%%%%%%%%%%%%%%%%%%%%%%%%%%%%%
\begin{figure}[h]
\includegraphics[draft=false,scale=0.6]{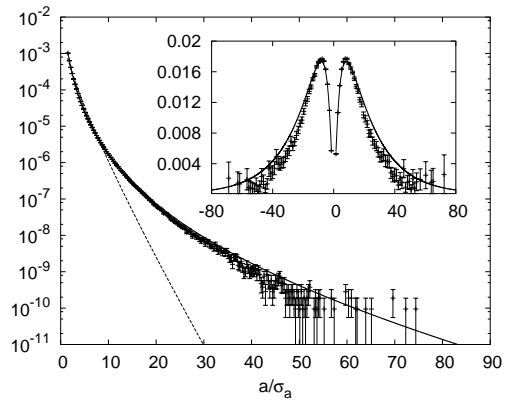}
\caption{Log-linear plot of the acceleration pdf. The crosses
are the DNS data, the solid line is the multifractal prediction and
the dashed line is the K41 prediction. 
The DNS statistics were calculated along the trajectories of two million particles 
amounting to $3.6 \times 10^9$ events in total. The statistical 
uncertainty in the pdf was quantified by assuming that fluctuations grow 
like the square root of the number of events. 
Inset: ${\tilde a}^4 {\cal P}({\tilde a})$ for the DNS data (crosses) and the
multifractal prediction.}
\label{fig:acc}
\end{figure}
%%%%%%%%%%%%%%%%%%%%%%%%%%%%%%%%%%%%%%%%%%%%%%%%%%%%%%%%%%%%%%%
In Fig.~(\ref{fig:acc}) we compare the acceleration pdf computed from
the DNS data with the multifractal prediction (\ref{pdf_norm}).  The
large number of Lagrangian particles used in the DNS (see
\cite{biferale04} for details) allows us to detect events up to $80$
$\sigma_a$. The accuracy of the statistics is improved by averaging
over the total duration of the simulation and all directions since the
flow is stationary and isotropic at small scales.  Also shown in
Fig.~(\ref{fig:acc}) is the K41 prediction for the acceleration pdf
${\cal P}^{K41}(\tilde{a}) \sim \tilde{a}^{-5/9} R_{\lambda}^{-1/2}
\exp \left( -\tilde{a}^{8/9}/2 \right)$ which can be recovered from (\ref{pdf_norm}) with
$h=1/3$, $D(h)=3$ and $\chi^{K41}=1$. 
As is evident from
Fig.~(\ref{fig:acc}), the multifractal  prediction (\ref{pdf_norm})
captures the shape of the acceleration pdf much better than the K41
prediction. What is remarkable is that (\ref{pdf_norm})  agrees
with the DNS data well into the tails of the distribution -- from the
order of one standard deviation $\sigma_a$ up to order $70 \sigma_a$.
This result is obtained with $D(h)$ given by (\ref{eq:d_h}).
We emphasise that the only degree of freedom in our formulation of ${\cal
P}(\tilde{a})$ is the minimum value of the acceleration, ${\tilde
a}_{\min}$, here taken to be $1.5$.  In the inset of Fig.~(\ref{fig:acc}) we
make a more stringent test of the multifractal prediction
(\ref{pdf_norm}) by plotting  ${\tilde a}^4 {\cal P}({\tilde a})$ and which is
seen to agree well with the DNS data. 

From (\ref{acc}) it is also possible to derive a prediction for the
acceleration moments conditioned on the local -- instantaneous --
velocity field $v_0$: $\langle a^n|v_0\rangle$. For example, for the
conditional acceleration variance we get 
$$
\left<a^2|v_0\right> \sim \int_{h \in I} \dd h \, 
\nu^{\frac{1+4h-D(h)}{1+h}} 
v_0^{\frac{3+D(h)}{1+h}} 
L_0^{\frac{D(h)-6h-3}{1+h}}.
$$
In the limit $\nu \ll 1$, a saddle point approximation gives 
$\left<a^2|v_0\right> \propto \nu^{\alpha} v_0^{(3+D(\hat{h}))/(1+\hat{h})}$
where $\alpha = \inf_h( (1 + 4h - D(h))/(1 + h) )$
and $\hat{h}$ is the value of $h$ which minimises the exponent of $\nu$.
Thus, we find that $\left<a^2|v_0\right> \propto v_0^{4.57}$.

%%%%%%%%%%%%%%%%%%%%a%%%%%%%%%%%%%%%%%%%%%%%%%%%%%%%%%%%%%%%%%%%
\begin{figure}[h]
\includegraphics[draft=false,scale=0.6]{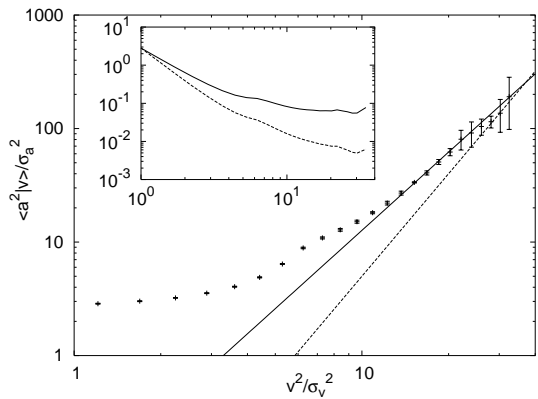}
\caption{Log-log plot of the conditional acceleration variance. 
The crosses are the DNS data, the solid line is the multifractal
prediction and the dashed line is the prediction of \cite{sawford}.
The DNS data is the absolute acceleration conditioned on the absolute
velocity.  Statistical uncertainty was estimated by dividing the
samples into five sub-ensembles.
Inset: the conditional acceleration variance scaled by
the multifractal prediction, $\langle a^2|v\rangle/v^{4.57}$,  
and the prediction of \cite{sawford}, $\left<a^2|v\right>/v^6$, 
both normalised by $\sigma_a^2$.}
\label{fig:cond-acc}
\end{figure}
%%%%%%%%%%%%%%%%%%%%%%%%%%%%%%%%%%%%%%%%%%%%%%%%%%%%%%%%%%%%%%%
In Fig.~\ref{fig:cond-acc} we plot $\langle a^2|v \rangle$, normalised
by the acceleration variance, versus $v^2/\sigma_v^2$. The relatively
large error that can be seen in the DNS conditional acceleration
statistics for large values of $v^2/\sigma^2_v$ reflects the rarity of
these events. However, in agreement with \cite{sawford} a clear trend
is evident that, for large velocities, the acceleration magnitude
depends strongly on the magnitude of the velocity. (The vector
acceleration and velocity can easily be shown to be uncorrelated for
stationary turbulence \cite{sawford}.)  Also shown in
Fig.~\ref{fig:cond-acc} are the multifractal prediction and the
prediction of \cite{sawford} based on a dimensional argument
pertaining to the vorticity, namely that $\left<a^2|v\right> \propto
v^6$. (This can be recovered from the multifractal formalism with
$h=0$, corresponding to a singularity in $\delta_{\tau} v$, and
$D(h)=3$.) Although statistical noise prevents us from making a
convincing claim, the multifractal prediction appears to agree better
with the DNS data.

In conclusion, we have shown that the multifractal formalism predicts
a pdf for the unconditional acceleration which is in excellent
agreement with the DNS data. In deriving the form of the pdf we have
only assumed that the Lagrangian velocity increment is related to
the Eulerian velocity increment  by (\ref{eq:tau_l}) and
that the large scale fluctuating velocity is Gaussian. The
only adjustable parameter in our formulation is the value of ${\tilde
a}_{\min}$ which does not have a sensitive effect on the
results. 

The simulations were performed on the IBM-SP4 of Cineca (Bologna,
Italy). We thank in particular G.~Erbacci and C.~Cavazzoni for 
resource allocation and precious technical assistance.  We acknowledge support
from the EU under the contracts HPRN-CT-2002-00300 and HPRN-CT-2000-0162.


\begin{thebibliography}{99}

\bibitem{K65} R.H. Kraichnan, Phys. Fluids {\bf 8}, 575 (1965).

\bibitem{pope} S.B. Pope, {\it Turbulent Flows}, 
(Cambridge University Press, Cambridge, 2000).

\bibitem{beck} C. Beck, Europhys. Lett., {\bf 64}, 151 (2003). 
C. Beck, Phys. Lett. A, {\bf 27}, 240 (2001). 

\bibitem{aringazin} A.K. Aringazin and M.I. Mazhitov, Phys. Lett. A, {\bf 313}, 284 (2003).

\bibitem{arimitsu} T. Arimitsu and N. Arimitsu, arXiv:cond-mat/0301516.

\bibitem{GK03} T. Gotoh and R.H. Kraichnan, arXiv:nlin.CD/0305040. 

\bibitem{cornell} A. La Porta {\it et al.}, Nature, {\bf 409}, 1017 (2001). 
G.A. Voth {\it et al}, J. Fluid Mech., {\bf 469}, 121 (2002). N. Mordant 
{\it et al.}, arXiv:physics/0303003.

\bibitem{pinton} N. Mordant {\it et al.}, J. Stat. Phys., {\bf 113}, 701 (2003). 
N. Mordant {\it et al.}, Phys. Rev. Lett., {\bf 89}, 254502 (2002). 
N. Mordant {\it et al.}, Phys. Rev. Lett., {\bf 87}, 214501 (2001). 

\bibitem{ott_mann} S. Ott and J. Mann, J. Fluid Mech., {\bf 422}, 207 (2000).

\bibitem{yeung} P.K. Yeung, Ann. Rev. Fluid Mech. {\bf 34}, 115 (2002).
P.K. Yeung, J. Fluid Mech. {\bf 427}, 241 (2001).
P. Vedula and P.K. Yeung, Phys. Fluids {\bf 11}, 1208 (1999).

\bibitem{BS02} G. Boffetta and I.M. Sokolov, Phys. Rev. Lett. {\bf 88}, 094501 (2002).

\bibitem{IK02} T. Ishihara and Y. Kaneda, Phys. Fluids {\bf 14}, L69 (2002).

\bibitem{GF01} T. Gotoh and D. Fukuyama, Phys. Rev. Lett. {\bf 86}, 3775 (2001). 

\bibitem{biferale04} L. Biferale {\it et al.}, arXiv:nlin.CD/0402032. 

\bibitem{frisch} U. Frisch, {\it Turbulence: the legacy of A.N. Kolmogorov}, 
(Cambridge University Press, Cambridge, 1995). 

\bibitem{parisi_frisch} G. Parisi and U. Frisch, in {\it Turbulence and predictability 
in geophysical fluid dynamics}, (North-Holland, New York, 1985), p. 84. 

\bibitem{benzi84} R. Benzi {\it et al.}, J. Phys., {\bf A17}, 3521 (1984). 

\bibitem{ben91} R. Benzi {\it et al.}, Phys. Rev. Lett., {\bf  67} 2299 (1991).

\bibitem{biferale93} L. Biferale, Phys. Fluids A, {\bf 5}, 428 (1993). 

\bibitem{borgas93} M.S. Borgas, Phil. Trans. R. Soc. Lond. A, {\bf 342}, 379 (1993).

\bibitem{chevillard} L. Chevillard {\it et al.}, Phys. Rev. Lett., {\bf 91}, 214502 (2003). 

\bibitem{BDM02} G. Boffetta {\it et al.}, Phys. Rev. E {\bf 66}, 066307 (2002).

\bibitem{she_leveque} Z.S. She and E. L\'ev\^eque, Phys. Rev. Lett., {\bf 72}, 336 (1994). 
 
\bibitem{benzi93} R. Benzi {\it et al.}, Phys. Rev. E, {\bf 48}, R29 (1993). 

\bibitem{sawford} B.L. Sawford {\it et al.}, Phys. Fluids {\bf 15}, 3478 (2003).

\bibitem{hill} R.J. Hill, J. Fluid Mech., {\bf 452}, 361 (2002). 
  
\end{thebibliography}
\end{document}